\documentstyle[preprint,aps,epsf]{revtex}
\begin{document}

\title{Extended bound states of fermions on 2D square lattice beyond the 
nn hopping and interactions }
 \author{M. B\c{a}k\footnote{karen@delta.amu.edu.pl} 
        and R. Micnas\footnote{rom@alpha.amu.edu.pl}}
\address {Institute of Physics, A. Mickiewicz University,\\ Umultowska 85, PL-61-614 Pozna\'n 
-- POLAND} 
%\date{\today }
\maketitle

\begin{abstract}
We examine extended bound states in the dilute limit of the extended Hubbard model with 
next-nearest-neighbour (nnn)
hopping and nnn interaction on the two dimensional square lattice. 
By the exact solution of a two-body problem we show that $d$-wave pairing is
strongly
favored by the nnn hopping. We determine the binding energies, mobilities 
and dispersion curves across Brillouin zone (B.z.) for bound states
of various symmetries.  The numerical analysis shows that the deep bound states can exist in the whole
B.z. and we have also found $s^{*}-d$ mixing of the bound
states.

\end{abstract}
\vspace{1cm}

\section{Introduction.}
The question of the symmetry of the superconducting order parameter 
arose very soon after the discovery of high temperature superconductivity (HTS).
 Different symmetries 
imply superconductivity of considerably different properties and large efforts have been
devoted to determining the type of pairing in different materials. Experimental findings have
yielded ambiguous results, yet they strongly suggest explanations extending beyond 
the simplest $s$-wave picture, 
possibly $d$-wave or mixed state in cuprate HTS. 
Recently, the angle-resolved photoemission spectroscopy and
tri-cristal Josephson effect\cite{shen}
 revealed a large $d$-wave component in the order parameter of cuprate
superconductors. 
Apart from different pairing mechanisms favoring this kind of pairing,
this situation may be due to kinetic mechanisms, as suggested by Blaer { et al}\cite{blaer}.
With inclusion of the nnn hopping, a bound pair formed on
adjacent lattice sites can propagate over long distances
before decaying. The effective mass of such a pair can become even smaller than
the mass of its constituent fermions. The issue of nnn hopping has also
important implications for the problem of hole motion in an antiferromagnetic background 
\cite{avracham} and  for the explanation of the HTS properties 
\cite{benard}--\cite{rmetal}. 

Exactly solvable model of two fermions on a periodic lattice would give a valuable insight 
into the
problems of both: the role of nnn hopping and the symmetry of the superconducting state.
It is also of importance, as a limiting case for the two relevant
 models of superconductivity: BCS and  preformed pairs scenario. 

In this paper we report results of exact calculations of bound states (BS) of 
fermions on a 2D square lattice including nnn hopping and nnn interactions,
using formulation of Ref. \cite{blaer}.
We have evaluated the pair binding energies, pair mobilities and dispersion curves across
the Brillouin zone for the extended Hubbard model for different pairing symmetries. 
We discuss in detail the 
effect of nnn hopping $t_{2}$ 
and nnn interactions $W_{2}$, on the stability of the shallow BS.

\section{Formalism}
\subsection{The model}
The details of the theoretical treatment are given in \cite{blaer} so we shall give here 
only the
main points. We start with the generalized Hubbard Hamiltonian on a periodic lattice with N sites:
\begin{eqnarray} \label{ham}
H=H_{0}+H_{1},\\
H_{0}=\sum_{\vec{r},\vec{r}\,',\sigma}t(\vec{r}-\vec{r}\,')c^{\dagger}_{\sigma}(\vec{r}\,')
c_{\sigma}(\vec{r}), \\ 
H_{1}=\sum_{\vec{r},\vec{r}\,'}[g_{S}(\vec{r}-\vec{r}\,')S^{\dagger}(\vec{r},\vec{r}\,')
S(\vec{r},\vec{r}\,')+
g_{T}(\vec{r}-\vec{r}\,')\sum_{\mu}T_{\mu}^{\dagger}(\vec{r},\vec{r}\,')
     T_{\mu}(\vec{r},\vec{r}\,')],
\end{eqnarray}
where $\vec{r}$, $\vec{r}\,'$ denote the positions of electrons and $S$, $T$ are singlet
and triplet pairing operators:
\begin{eqnarray}
S(\vec{r},\vec{r}\,')=
    \frac{c_{\uparrow}(\vec{r})c_{\downarrow}(\vec{r}\,') + c_{\uparrow}(\vec{r}\,')
c_{\downarrow}(\vec{r})}{2},\\
T_{0}(\vec{r},\vec{r}\,')=\frac{c_{\uparrow}(\vec{r})c_{\downarrow}(\vec{r}\,') -
c_{\uparrow}(\vec{r}\,')c_{\downarrow}(\vec{r})}{2},\\
T_{+}(\vec{r},\vec{r}\,')=\frac{c_{\uparrow}(\vec{r})c_{\uparrow}(\vec{r}\,')}{\sqrt{2}},\\
T_{-}(\vec{r},\vec{r}\,')=\frac{c_{\downarrow}(\vec{r})c_{\downarrow}(\vec{r}\,')}{\sqrt{2}}.
\end{eqnarray}
In the standard Hubbard model notation $g_{S}(R_{i},R_{i})=U$, $g_{S}(R_{i},R_{j})=W_{ij}$ and
$g_{T}(R_{i},R_{j})=W_{ij}$. The inclusion of antiferromagnetic exchange
$\sum_{ij}J_{ij}\!\cdot\!\vec{S_{i}}\!\cdot\!\vec{S_{j}}$, $J_{ij}>0$ amounts to change
$g_{S}(R_{i},R_{j})=W_{ij}-3J_{ij}/2$, $g_{T}(R_{i},R_{j})=W_{ij}+J_{ij}/2$
and triplet pairing can be suppressed. From now on we will consider the singlet pairing
only.

A convenient basis set 
in the Hilbert space of two fermions
consists of states 
$|\vec{P},\vec{r},\sigma_{1},\sigma_{2}>$ 
with total Bloch momentum $\vec{P}$ and relative separation $\vec{r}$. In relative
coordinates in the reciprocal space 
the eigenvectors of the kinetic part of the Hamiltonian (\ref{ham}) are the states:
\begin{equation}
|\vec{P},\vec{p},\sigma_{1},\sigma_{2}>=
c^{\dagger}_{(\vec{P}/2)-\vec{p},\sigma_{1}}c^{\dagger}_{(\vec{P}/2)+\vec{p},\sigma_{2}}|0>,
\end{equation}
with the eigenvalues:
\begin{equation}
E_{\vec{P}\vec{p}}=\epsilon_{\vec{P}/2-\vec{p}}+\epsilon_{\vec{P}/2+\vec{p}}\;,
\end{equation}
where 
\begin{equation}\label{ekk}
\epsilon_{\vec{p}}=-2t_{1}(\cos(p_{x}\delta)+\cos(p_{y}\delta))
-4t_{2}\cos(p_{x}\delta)*\cos(p_{y}\delta).
\end{equation}
where $\delta$ is a lattice constant.  We also define
\begin{eqnarray}
\frac{1}{\sqrt{N}}\psi(\vec{r}|\vec{p})=<\vec{P},\vec{r},\sigma_{1},
            \sigma_{2}|\vec{P},\vec{p},\sigma_{1},\sigma_{2}>=e^{i\vec{p}\cdot\vec{r}}, \\
\frac{1}{\sqrt{N}}\psi(\vec{r})=<\vec{P},\vec{r},\sigma_{1},\sigma_{2}|
                    \vec{P},\psi,\sigma_{1},\sigma_{2}>,
\end{eqnarray}
where $\psi(\vec{r})$ is a wave function in  the relative position space, 
a result of projection of any two-fermion state on a state with definite relative 
separation $\vec{r}$ 
and total Bloch momentum $\vec{P}$.
To solve the Schr\"{o}dinger equation
\begin{equation}
(H_{0}+H_{1})|\vec{P},\psi>=E|\vec{P},\psi>,
\end{equation}
where $H_{0}$ is the kinetic part of the Hamiltonian (\ref{ham}), we introduce:
\begin{equation}
|\vec{P},\psi>=\sum_{p}F(\vec{p})|\vec{P},\vec{p}>,
\end{equation}
which yields an integral equation for coefficients $F(p)$:
\begin{eqnarray}\\ \nonumber
(E-E_{\vec{P}\vec{p}})F(\vec{p})=\sum_{\vec{q}}F(q)<\vec{P}\vec{p}|H_{1}|\vec{P}\vec{q}>=\\ 
\frac{1}{N}\sum_{\vec{r}}\psi^{*}(\vec{r}|\vec{p})g(\vec{r})
                \sum_{\vec{q}}F(\vec{q})\psi(\vec{r}|\vec{q}).
\end{eqnarray}

Using the above formula for $F(p)$ and definition:
$\psi(\vec{r})=\sum_{\vec{p}}F(\vec{p})\psi(\vec{r}|\vec{p}),$
one obtains the final equation for the wave function $\psi(\vec{r})$:
\begin{equation}\label{psi1}
\psi(\vec{r})=\sum_{\vec{r}\,'}G(E,\vec{p},\vec{r},\vec{r}\,')g(\vec{r}\,')\psi(\vec{r}\,'),
\end{equation}
where:
\begin{equation}
G(E,\vec{P},\vec{r},\vec{r}\,')=\frac{1}{N}\sum_{\vec{q}}\frac{\psi(\vec{r}|\vec{q})
     \psi^{*}(\vec{r}\,'|\vec{q})}{E-E_{\vec{P}\vec{q}}},
\end{equation}
is the Green function.

For the sites where interaction $g$ is present, Eq. (\ref{psi1}) can be cast into the form:
\begin{equation}\label{main}
[1-{\cal G}(E,\vec{P})g]\psi=0,
\end{equation}
where $\psi$ is a $n \times 1$ matrix with $\psi_{j}=\psi(R_{j})$, ${\cal G}(E,\vec{P})$ is a
$n \times n$ matrix with ${\cal G}_{ij}(E,\vec{P})=
G(E,\vec{P},R_{i},R_{j})$ and $g$ is a diagonal
$n \times n$ matrix with $g_{ii}=g(R_{i})$ where $g(R_{i})\neq 0$ only
for sites $\vec{r}=0,\pm R_{1},...,\pm R_{n-1}$. 
The energy eigenvalues are calculated from the determinant:
\begin{equation}\label{det}
det[1-{\cal G}(E,\vec{P})g]=0.
\end{equation}
To get the wave function for lattice sites outside described range we 
have to use Eq. (\ref{psi1}).

\subsection{Final equations}
The matrix element in the reciprocal space is given as:
\begin{equation}\label{mx}
N<\vec{P}\vec{p}|H_{1}|\vec{P}\vec{q}>=
g_{0}+g_{1}\gamma^{(1)}_{\vec{p}-\vec{q}}+g_{2}\gamma^{(2)}_{\vec{p}-\vec{q}}\;,
\end{equation}
where: 
$\gamma^{(1)}_{\vec{p}}=2[\cos(p_{x}\delta)+\cos(p_{y}\delta)]$,
$\gamma^{(2)}_{\vec{p}}=4\cos(p_{x}\delta)\cos(p_{y}\delta)$,
$g_{0}=U$ is the on-site interaction, $g_{1}=W_{1}$ and $g_{2}=W_{2}$ 
denote intersite interactions between nn and nnn, respectively.
A separation of the matrix element (\ref{mx}) into parts depending exclusively on $\vec{p}$ and 
$\vec{q}$ 
is done by using five components $\psi_{n}(\vec{p})$, transforming according to
irreducible representations of the symmetry group of the $d=2$ lattice $C_{4v}$ 
(given in the third column):
\begin{equation}
\begin{array}{ccc}
 \psi_{0}=1,                                  & \;\;\;\;\;\; \mbox{($s$-wave)} &
A_{1}\\ 
 \psi_{1}(\vec{p})=cos(p_{x}\delta)+cos(p_{y}\delta),  & \;\;\;\;\;\; \mbox{(ext. $s$*-wave)} &
A_{1}\\ 
 \psi_{2}(\vec{p})=cos(p_{x}\delta)-cos(p_{y}\delta),  & \;\;\;\;\;\; \mbox{($d_{x^{2}-y^{2}}$-wave)} &
B_{1}\\
 \psi_{3}(\vec{p})=2cos(p_{x}\delta)*cos(p_{y}\delta), & \;\;\;\;\;\; \mbox{($s_{xy}$*-wave)} &
A_{1}\\ 
 \psi_{4}(\vec{p})=2sin(p_{x}\delta)*sin(p_{y}\delta). & \;\;\;\;\;\; \mbox{($d_{xy}$-wave)}
& B_{2}
\end{array}
\end{equation}

With such a form of the basis functions the Green's function is given by:
\begin{equation}\label{gie2}
G_{nm}(E,\vec{P})=\frac{1}{N}\sum_{\vec{p}}\frac{\psi_{n}(\vec{p})\psi_{m}(\vec{p})}
      {E-E_{\vec{P}\vec{p}}}.
\end{equation}
Above decomposition allows to write the following equation for $F(\vec{p})$
being the Fourier transform of $\psi(\vec{r})$:
\begin{equation}
F_{m}=\sum_{n\in\cal{R}}g_{n}G_{nm}(E,\vec{P})F_{n},
\end{equation}
for each irreducible representation,
 where $F_{m}=(1/N)\sum_{\vec{p}}F(\vec{p})\psi_{m}(\vec{p})$. 
Equation (\ref{det}) for eigenvalues $E$
takes the form shown below --
for $s^{*}$-wave pairing
(compare also \cite{teubel,rmetal}): 
\begin{equation}\label{mac}
det \left( \begin{array}{ccc}
1-G_{00} U & -G_{01} W_{1}   & -G_{03} W_{2} \\
-G_{01} U   & 1-G_{11} W_{1} & -G_{13} W_{2} \\
-G_{03} U   & -G_{13} W_{1}   & 1-G_{33} W_{2}
\end{array} \right) = 0,
\end{equation}
for $d_{x^{2}-y^{2}}$-wave:
\begin{equation}\label{mac+1}
1+|W_{1}|G_{22}=0,
\end{equation}
and for $d_{xy}$:
\begin{equation}
1+|W_{2}|G_{44}=0.
\end{equation}

The above decomposition holds for arbitrary $t_2$ and $\vec{P}=0$ or $t_2=0$ and $\vec{P}$ on the $\Gamma M$ line as well as for any $t_2$ with $\vec{P}$ on the $\Gamma M$ line and $W_2=0$.
In general case one has to resort to Eq.(\ref{det}).

Eq. (\ref{mac}) reads:
\begin{eqnarray}\label{finalS}
(1-G_{00}U)(1-G_{11}W_{1})(1-G_{33}W_{2}) - 
U W_{1} W_{2}(2G_{01}G_{13}G_{03}-G_{03}^{2}G_{11}-\nonumber \\
G_{13}^{2}G_{00}-G_{01}^{2}G_{33})-G_{03}^{2}UW_{2}-G_{13}^{2}W_{1}W_{2}-G_{01}^{2}UW_{1}=0.
\end{eqnarray}
From Eq. (\ref{finalS}) we can derive simpler cases\cite{rmetal}:
\begin{equation}
\begin{array}{cl}
 G_{11}=-1/|W_{1}|,                         & \;\;\;\;\;\;\;\mbox{for $W_{2}=0=U$;}\\
      \label{last}
 -G_{00}(1+G_{11}|W_{1}|)+G_{01}^{2}|W_{1}|=0, & \;\;\;\;\;\;\;\mbox{for $W_{2}=0,U=\infty$.}
\end{array}
\end{equation}

The relative pair mobility which is defined as:
\begin{equation}
Z=\frac{2m_{f}}{m^{*}}
\end{equation}
where $m^{*}$ is an effective mass of a bound pair, and
for $a\equiv t_{2}/t_{1}\neq 0$, 
for two-dimensional square lattice $2m_{f}=\hbar^{2}/(t_{1}+2t_{2})\delta^{2}$.
Effective mass $m^{*}$ can be obtained by expanding the denominator in Eq.(\ref{gie2}). 
We take:
\begin{equation}
E-E_{\vec{P}\vec{p}}=2\epsilon_{0}-E_{b}-\omega_{\vec{P}}-\epsilon_{\vec{P}/2-\vec{p}}-
    \epsilon_{\vec{P}/2+\vec{p}}\;,
\end{equation}
where $E_{b}$ is the binding energy at $\vec{P}=0$, $\omega_{\vec{P}}$ 
is the
dispersion of the bound state and $\epsilon_{\vec{p}}$ is given by Eq. (\ref{ekk}).
In the long-wavelength limit on the $\Gamma M$ line: $\omega_{\vec{P}}=D P^{2}$ where $D=\hbar^{2}\delta^{2}/2m^{*}$.
The coefficient $D$ is obtained by expanding $\epsilon_{\vec{P}/2\pm \vec{p}}$ 
into powers of $\vec{P}$, inverting the series and substituting resulting
$G_{nm}(E,\vec{P})$ into appropriate Eqs (\ref{mac+1} -- \ref{last}). The final
equations for pair mobilities in the simplest cases are:
\begin{equation}
Z=\frac{1}{1+2a}*\frac{A_{nn}}{B_{nn}},
\end{equation}
where $n=1$ for $s^{*}$-wave pairing with $U=0$ and $W_{2}=0$; 
$n=2$ for $d_{x^{2}-y^{2}}$ and $n=4$ for
$d_{xy}$ pairing with $a=0$, respectively.
We have used denotations:
\begin{eqnarray}
A_{nm}=\frac{1}{2} B_{nm1} + a B_{nm3},\\
B_{nms}=\frac{1}{N}\sum_{p}\frac{\psi_{n}(p)\psi_{m}(p)\psi_{s}(p)}
         {(2\epsilon_{0}-E_{b}-2\epsilon_{p})^{2}},\\
B_{nm}=B_{nm0}.
\end{eqnarray}

Mobilities in other cases have more complicated forms.

\section{Results}
In Figs (1) - (3) we have plotted the binding energies and mobilities vs $|W_{1}/8t_{1}|$ 
for extended-$s$ pairing
(denoted as $s^{*}$) and for $d_{x^{2}-y^{2}}$-wave
pairing (in short $d$), for nnn
hopping with parameter $a=0$, $a=-0.2$ and $a=-0.45$. The last two values roughly correspond to 
the values of 
the parameter $a$, suggested by the band structure calculations in $La_{2-x}Sr_{x}CuO_{4}$ 
($a\simeq -0.25$) and $YBa_{2}Cu_{3}O_{7-x}$ ($a\simeq
-0.45$) \cite{benard}.

With only the nn hopping taken into account (i.e. $a=0$) for $U=0$ 
infinitesimally small
value of $|W_{1}|$ can create a bound state (of the $s^{*}$ type) -- there is no
critical value. For finite $U$ or $W_{2}$, including the case of
the infinite on-site 
repulsion, $|W_{1}|$ has to reach critical value to create such a bound state
($|W_{1c}|/8t_{1}=0.25$ for $s^{*}$-wave for $U=\infty$ for $a=0$). 
The $d$-wave pairing
does not depend on $U$, and also a critical value of $|W_{1c}|/8t_{1}=0.915$ 
(or $|W_{2c}|/8t_{1}=0.8268$ in the case of
$d_{xy}$) has to be reached to form a pair of 
the $d$ type. In this case ($a=0$), the binding energies of $d$ type are smaller than 
the $s$-wave binding energies
in the whole range of displayed parameters. 
The mobility of pairs is smaller then 1 (i.e. the effective mass of a pair is
larger then the mass of its constituent fermions) and decreases with increasing 
intersite attraction.
Increasing $|W_{1}|$ makes the binding stronger, but at the same time, the
pair movement more difficult. 
The mobility increases
with decreasing $|W_{1}|$ for $a=0$, and is always smaller for $d$-wave than $s$-wave. 
Let us note, that the mobility of the $s^{*}$-pairs for $U=\infty$ becomes larger than the
mobility of the $s^{*}$-pairs for $U=0$ near the critical value of $W_{1}$. It
is so, despite the fact, that the binding energies for $U=\infty$ are smaller than
in the $U=0$ case. It is also of interest, that both mobility and binding energies of
the $s^{*}$-pairs for $U=\infty=W_{2}$ resemble the respective quantities of the
$d$-wave pairing. Differences appear near the critical value, which is
slightly smaller in the $s^{*}$ case, with $s^{*}$ mobility there about twice as large as in
the $d$ case, but they diminish very quickly with increasing $|W_{1}|$. 

The situation changes
rapidly with increasing $t_{2}$ (which is taken with the opposite sign than $t_{1}$). 
The binding energy of $d$-wave pairs  
grows and can become larger than that of $s$-wave pairs for large enough $|W_{1}|$.
The same happens for the mobilities.
For the $d$-wave, the mobility grows faster than $s$-wave and for large enough $|W_{1}|$ 
(although not the
same as for pair binding energies) the mobility 
of $d$-pairs exceeds that of $s$-pairs.
Starting from $|a|\approx 0.3$ the pair mobilities can become greater than 1 and
even strongly bound pair can easily move without breaking its bond. 
For $|W_{1}|\rightarrow\infty$,
$Z\rightarrow \pm |a|/(1+2a)$ with $+$($-$) for $d$ ($s^{*}$) -wave
pairs, respectively\cite{blaer}.

The  increase of pair mobilities with the 
parameter $|a|$ is much more abrupt for $d$-wave pairing than
for $s$-wave, so at the value $a=-0.45$ the maximum of $d$-pairs mobility is around 6 while that 
of $s$-pairs around 4, and  the 
mobility for $s$-wave is smaller than for $d$-wave in the whole range of displayed
$W_{1}$ values (Fig. 3). 

With growing $|a|$ a new feature develops: a maximum appears in the mobility curves. The 
mobility generally increases with $a$
and decreases with $|W_{1}|$ but is not the largest for the critical value of $W_{1}$.
For $s^{*}$-wave pairing with $U=0$ the mobility for $W_{1}=0$ is always equal to the
mass of two free fermions, i.e. 1 in our units. For small binding energies, the mobility
increases with $|W_{1}|$ and diminishes only later, for still larger $|W_{1}|$, which creates
a maximum. This is a consequence of the saddle points located at $(0,\pi)$ and
$(\pi,0)$  in the reciprocal
space. The dispersion curves are anomalously flat in that region (as can be
seen in Fig. 5),  particles move
slower and the probability of pairing increases. The singularity is only
"felt" by the shallow states, which results in a maximum for small $|W_{1}|$.

We note that for large enough $|W_{1}|$ the 
mobility of $s^{*}$ bound pairs becomes
negative, which is connected to the fact that the minimum of BS energy moves from $k=0$ 
towards $k=\pi$.

In Figs (4) and (5) we show the bound state dispersion curves across the Brillouin zone
for $W_{2}=0$.
The BS energies of $d$- and $s^{*}$- pairing are the same at
the zone boundary. At the $\Gamma$ point $d$-wave BS energy is larger or smaller 
than $s$-wave BS energy, depending on $a$ and $|W_{1}|$.
The $d$-wave BS energies can be larger
than $s$-wave ones in a part of the B. z. and smaller in another part of the B. z. and the
possibility of mixed $s^{*}-d$ state emerges.

In Fig. (5) we plotted the dispersion curves
along the lines $\Gamma - M - X - \Gamma$  for $a=-0.45$. 
We observe a mixing of $s^{*}$ and $d$ wave states and well developed dispersion for deep bound states.
Calculations for small binding energies show, however, that shallow bound states may
exist only in parts of the B.z. We were unable to to find bound states in the
domain of real energies near the $M$ point 
for small $|W_{1}|$ (both in the case of $s^{*}$- and $d$-wave
pairing).
There is a possibility  that BS enter the
scattering continuum and turn into resonance states.
The problem of the resonance states is left for further study.

Fig. (6) shows the ground state diagram for the shallow bound states.  
The nnn interaction $W_{2}$ was assumed to be positive, i.e. we 
disregarded $d_{xy}$ pairing.
The resulting curves show, that for
nn hopping $s^{*}$-wave pairing prevails even for the infinite 
nnn repulsion.  The inclusion of $t_{2}$ together with the 
on-site repulsion act in favour of $d$-wave pairing.
For a given $U$, a finite $W_{2}$ is needed to get the ground state of $d$-wave symmetry.

Finally, let us point out that our results for
existence of two-body BS in the $s$-wave channel are directly related to the
stability of $s$-wave superconductivity on 2D lattice in a dilute limit~\cite{randeiraDuan}.

\section{Conclusions}
We have examined two-electron bound states in an extended Hubbard model with nnn
hopping and nnn interaction. The analysis shows that $t_{2}$ affects kinetic
energy of pairs in a considerable way. This second hopping, with opposite sign to 
$t_{1}$, enhances $d$-wave pairing more
than $s$*-wave. The nnn hopping increases the relative pair mobility, 
which can become greater 
than that of free fermions (and a possibility of long-lived higly mobile bound pairs arises),
in agreement with \cite{blaer}.
The increase of $d_{x^{2}-y^{2}}$-wave pair mobility is much larger than the respective one of 
the $s$*-wave.
The nnn interaction influences the $d$-wave pairing 
only of the $d_{xy}$ type.
 We have also
exemplified the 
possibility of existence of a $s^{*}-d_{x^{2}-y^{2}}$ mixed state. We have also
 pointed out
that for large  $t_{2}$  shallow bound states may exist only in certain
regions of B.z., in contrast to the deep bound states which can occur in the
whole B.z., for strong enough intersite attraction.

\section{Acknowledgments}
This paper was supported by the State Committee for Scientific Research (KBN
Poland): Project No~2~PO3B~056~14.

\newpage
%1
\begin{figure}[h]
\begin{flushleft}
\leavevmode
\epsfxsize=11cm
\epsfbox{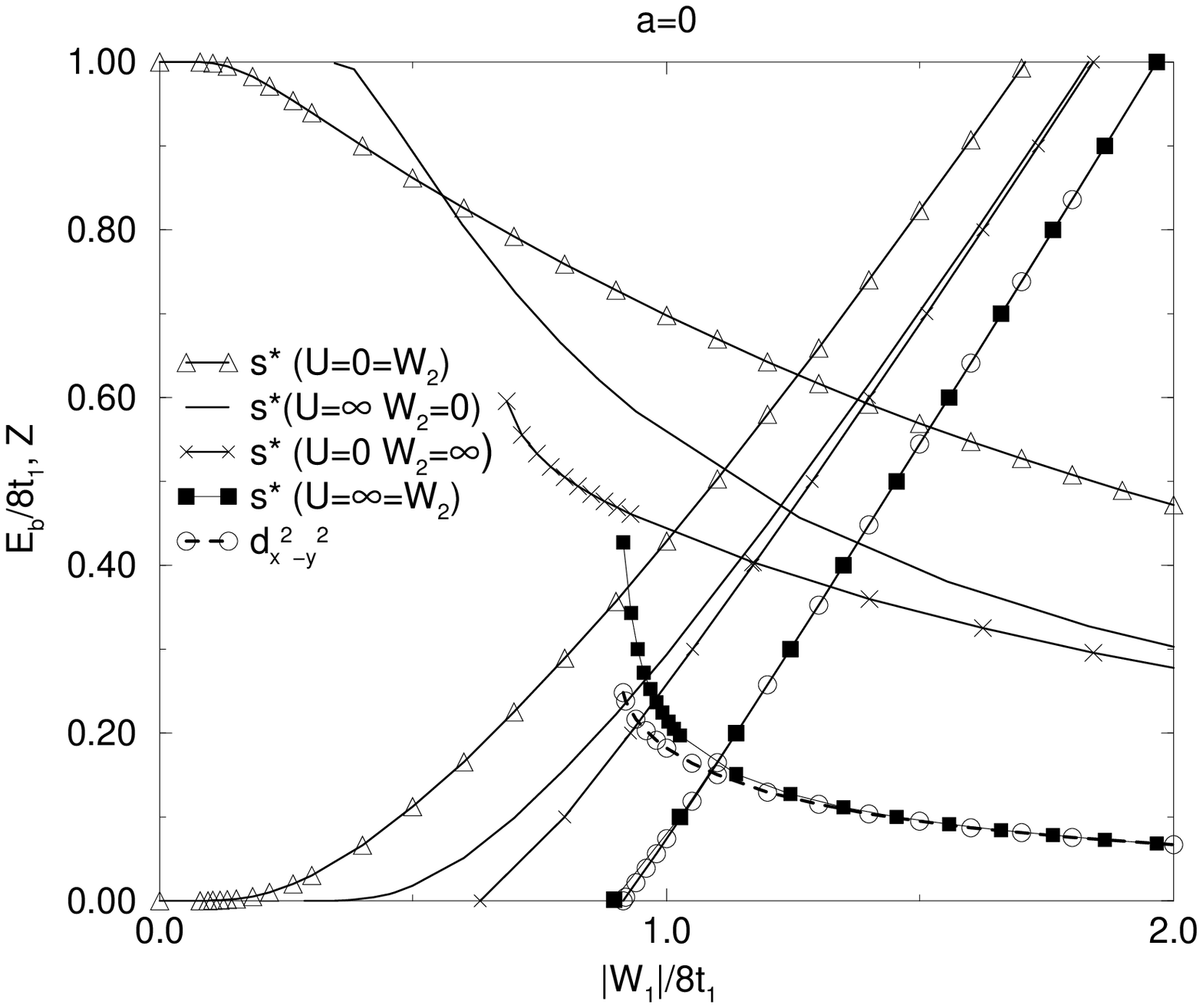}
\caption[]{Binding energies (curves increasing with $|W_{1}|$) and pair mobilities 
(curves decreasing with $|W_{1}|$)
on $d=2$ square lattice with nn hopping ($a=0$) 
for $s^{*}$-wave pairing: for $U=0$, $W_{2}=0$;
for $U=0$, $W_{2}=\infty$; for $U=\infty$, $W_{2}=0$; for $U=\infty$, $W_{2}=\infty$; and for
$d$-wave. }
\end{flushleft}
\end{figure}

%2
\begin{figure}[h]
\begin{flushleft}
\leavevmode
\epsfxsize=11cm
\epsfbox{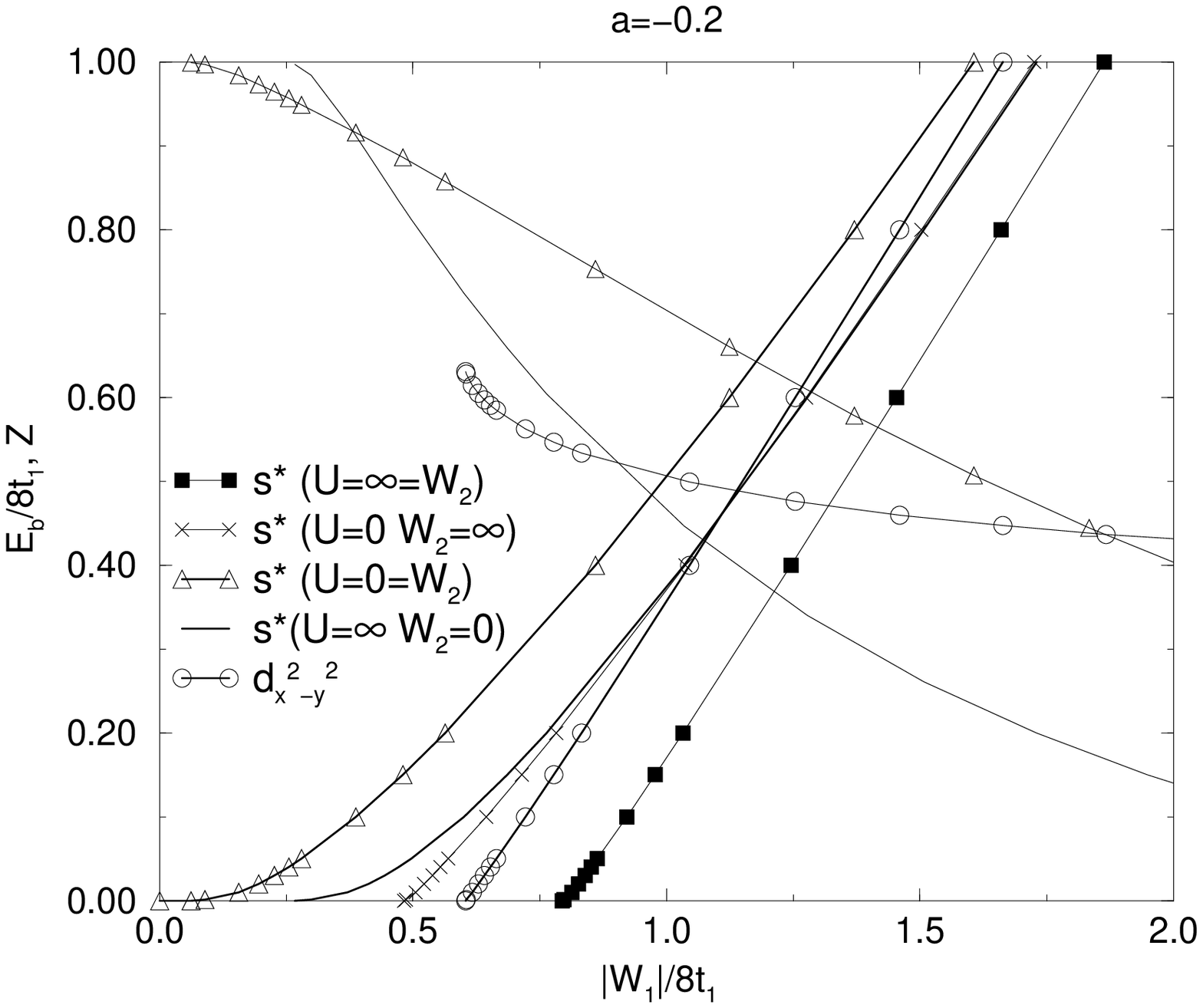}
\caption[]{Binding energies (curves increasing with $|W_{1}|$) and pair mobilities 
(curves decreasing with $|W_{1}|$)
on $d=2$ square lattice with nnn hopping ($a=-0.2$) 
for $s^{*}$-wave pairing: for $U=0$, $W_{2}=0$;
for $U=0$, $W_{2}=\infty$; for $U=\infty$, $W_{2}=0$; for $U=\infty$, $W_{2}=\infty$; and for
$d$-wave. }
\end{flushleft}
\end{figure}

%3
\begin{figure}[h]
\begin{flushleft}
\leavevmode
\epsfxsize=11cm
\epsfbox{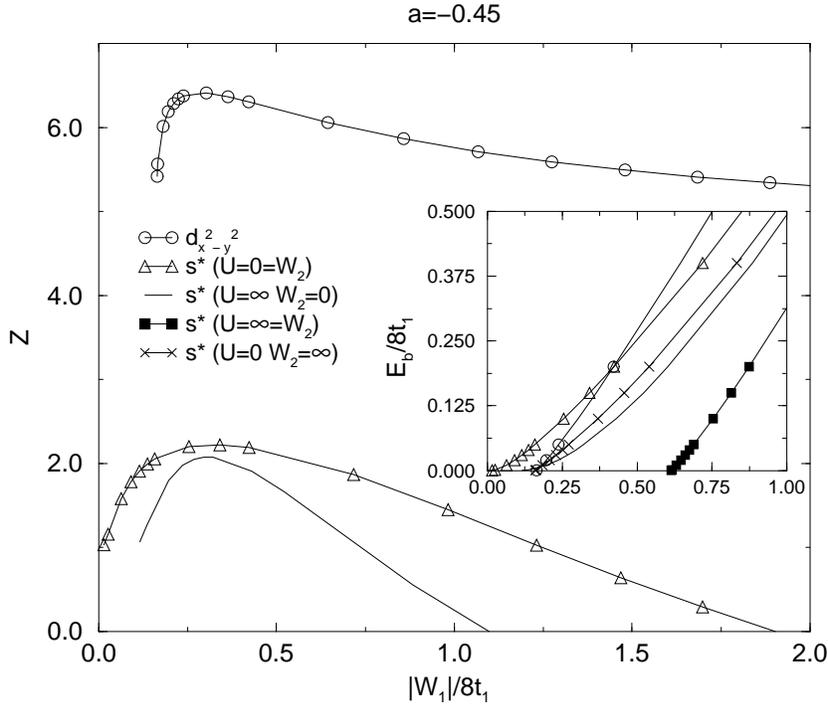}
\caption[]{Binding energies (inset) and pair mobilities for $a=-0.45$ for $s^{*}$-wave pairing for
$U=0=W_{2}$,
for $U=0$, $W_{2}=\infty$, for $U=\infty$, $W_{2}=0$, for $U=\infty=W_{2}$, and for
$d$-wave. }
\end{flushleft}
\end{figure}

%4
\begin{figure}[h]
\begin{flushleft}
\leavevmode
\epsfxsize=11cm
\epsfbox{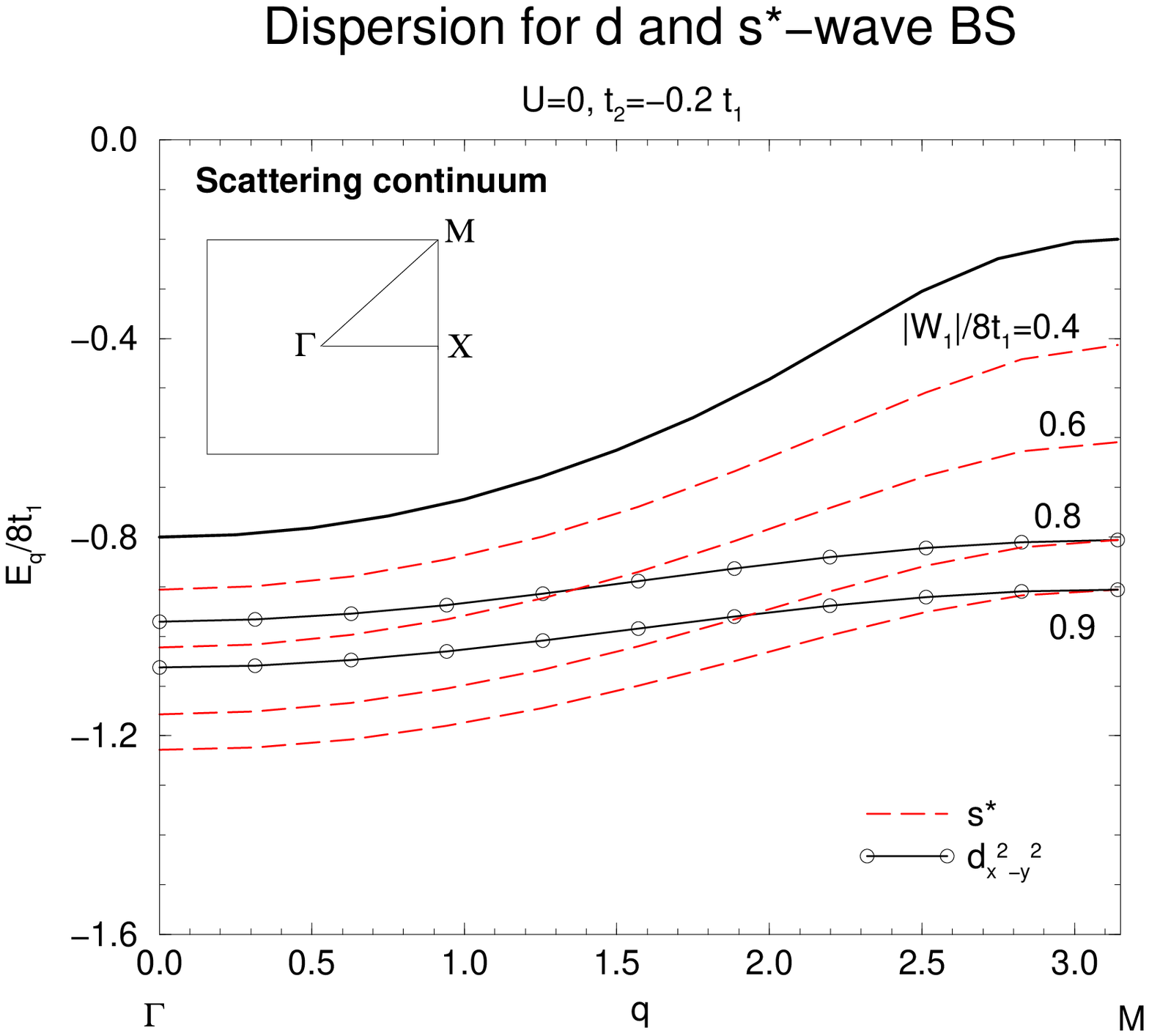}
\caption[]{Dispersion curves for $d$-wave and $s^{*}$-wave bound pairs for $a=-0.2$.
Heavy line denotes the lower boundary of scattering continuum: $min_{\vec{p}}\; E_{\vec{P}\vec{p}}$.
$q=P$ is the total momentum.}
\end{flushleft}
\end{figure}

%5
\begin{figure}[h]
\begin{flushleft}
\leavevmode
\epsfxsize=13cm
\epsfbox{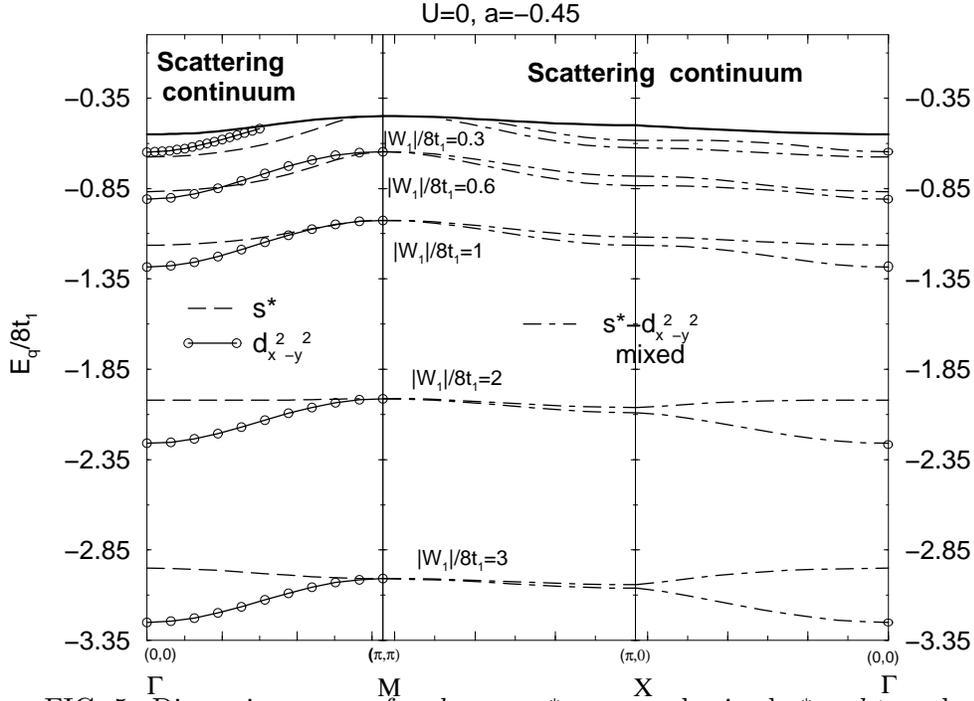}
\caption[]{Dispersion curves for $d$-wave, $s^{*}$-wave and mixed $s^*-d$ type
    bound pairs for $a=-0.45$ along
the lines $\Gamma-M-X-\Gamma$.}
\end{flushleft}
\end{figure}

%6
\begin{figure}[h]
\begin{flushleft}
\leavevmode
\epsfxsize=12cm
\epsfbox{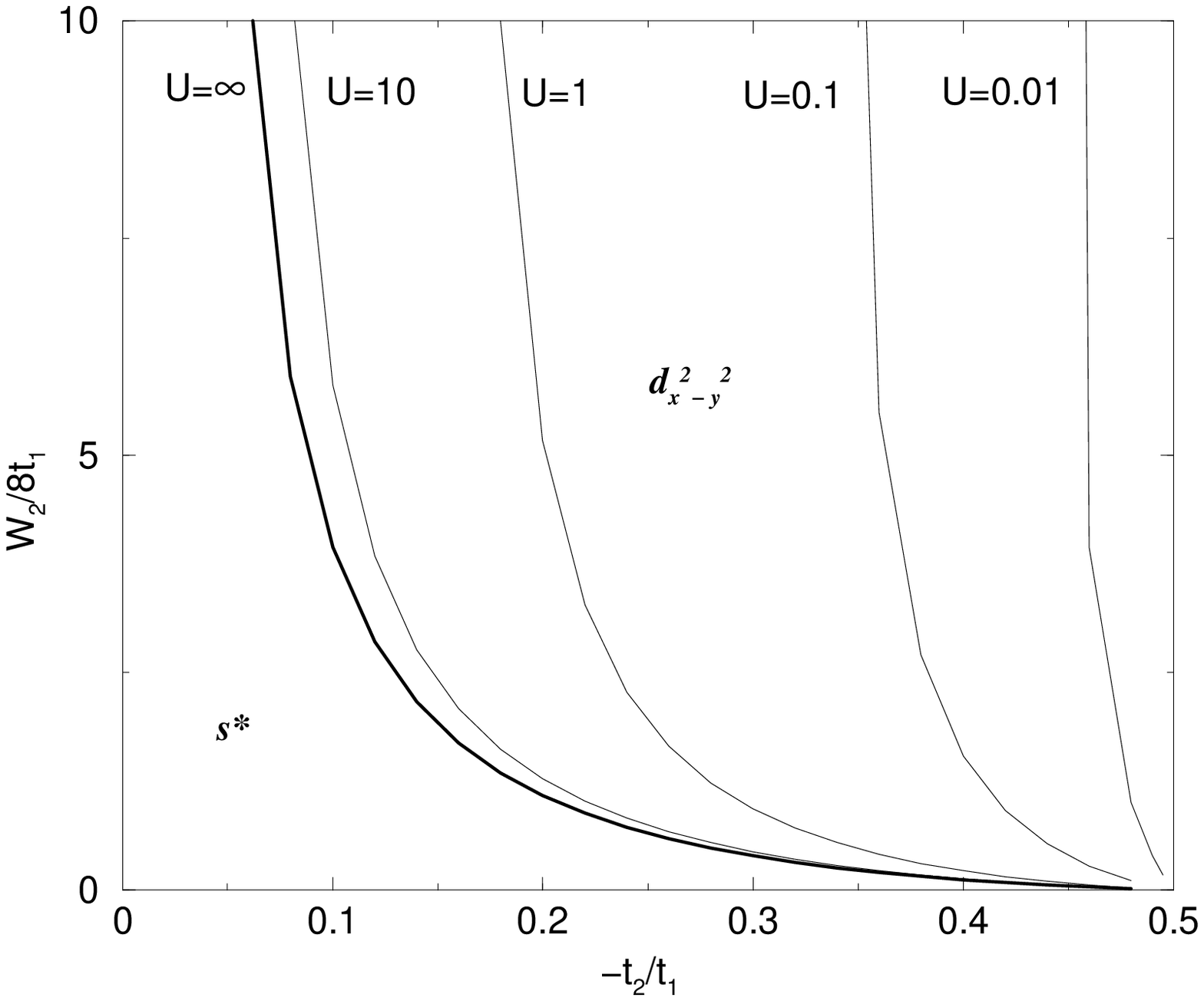}
\caption[]{Stability diagram for the shallow bound states for $W_{2}>0$. U in $8t_{1}$
units.}
\end{flushleft}
\end{figure}


\begin{references}
\bibitem{blaer} A. S. Blaer, H. C. Ren and O. Tchernyshyov, Phys. Rev. {\bf B 55}, 6035 (1997).
\bibitem{shen} Z. X. Shen { et al.}, Phys. Rev. Lett. {\bf 70}, 1553 (1993); 
  C. C. Tsuei { et al.}, Phys. Rev. Lett. {\bf 73}, 593 (1994).
\bibitem{avracham} A. Shiller { et al}, Phys. Rev. {\bf B 51}, 8337 (1995).
\bibitem{benard} P. B\'{e}nard, L. Chen and A. - M. S. Tremblay, Phys. Rev. {\bf B 47}, 
15217 (1993).
\bibitem{teubel} A. Teubel, E. Kolley and W. Kolley, phys. stat. sol. (b) {\bf 157}, 389 (1990).
\bibitem{rmetal} R. Micnas, J. Ranniger and S. Robaszkiewicz, Rev. Mod. Phys. 
   {\bf 62}, 112 (1990).
\bibitem{randeiraDuan} M. Randeira, J.-M. Duan, L.-Y. Shieh, Phys. Rev. B
{\bf 41} 327 (1990).
\end{references}
\end{document}